\documentclass[aps,prd,10pt,twocolumn,superscriptaddress,nofootinbib,showkeys,showpacs,altaffilletter]{revtex4-1}

\usepackage{graphicx,color}
\usepackage{rotating}
\usepackage{amssymb}
\usepackage{amsfonts}
\usepackage{amsmath}

\usepackage{lipsum}
\usepackage{booktabs,array}

\newcommand{\be}{\begin{equation}}
\newcommand{\ee}{\end{equation}}
\newcommand{\bea}{\begin{eqnarray}}
\newcommand{\eea}{\end{eqnarray}}

\date{\today}

\begin{document}


\title{Ricci cosmology in light of astronomical data} 

\author{Roberto Caroli}
 \email{roberto.caroli@usz.edu.pl}
\affiliation{Institute of Physics, University of Szczecin, Wielkopolska 15, 70-451 Szczecin, Poland}

\author{Mariusz P. D\c{a}browski}
\email{mariusz.dabrowski@usz.edu.pl}
 \affiliation{Institute of Physics, University of Szczecin, Wielkopolska 15, 70-451 Szczecin, Poland} 
 \affiliation{National Centre for Nuclear Research, Andrzeja Soltana 7, 05-400 Otwock, Poland}
 \affiliation{Copernicus Center for Interdisciplinary Studies, Szczepa\'{n}ska 1/5, 31-011 Krak\'{o}w, Poland}
\author{Vincenzo Salzano}
 \email{ vincenzo.salzano@usz.edu.pl}
\affiliation{Institute of Physics, University of Szczecin, Wielkopolska 15, 70-451 Szczecin, Poland}

\date{\today}

\begin{abstract}
Recently, a new cosmological framework, dubbed Ricci Cosmology, has been proposed. Such a framework has emerged from the study of relativistic dynamics of fluids out of equilibrium in a curved background and is characterised by the presence of deviations from the equilibrium pressure in the energy-momentum tensor which are due to linear terms in the Ricci scalar and the Ricci tensor. The coefficients in front of such terms are called the second order transport coefficients and they parametrise the fluid response to the pressure terms arising from the spacetime curvature. 

Under the preliminary assumption that the second order transport coefficients are constant, we find the simplest solution of Ricci cosmology in which the presence of pressure terms causes a departure from the perfect fluid redshift scaling for matter components filling the Universe. In order to test the viability of this solution, we make four different ans\"{a}tze on the transport coefficients, giving rise to four different cases of our model.  On the physical ground of the second law of thermodynamics for fluids with non-equilibrium pressure, we find some theoretical bounds (priors) on the  parameters of the models. Our main concern is then the check of each of the case  against the standard set of cosmological data in order to obtain the observational bounds on the second order transport coefficients. We find those bounds also realising that Ricci cosmology model is compatible with $\Lambda$CDM cosmology for all the ans\"{a}tze. 
\end{abstract}

\maketitle


\section{\label{sec:intro}Introduction\protect}

 The discovery of the late-time accelerated expansion of the Universe was made in the late 1990s, observing fainter than previously predicted Type Ia Supernovae (SnIa) \cite{Riess:1998cb,Perlmutter:1998np}. Since then, more and more precise observations, such as Baryon Acoustic Oscillation (BAO) \cite{Weinberg:2012es} and Cosmic Microwave Background (CMB) anisotropies \cite{Aghanim:2018eyx}, have pointed in the same direction making the explanation of these observations so compelling that it has become one of the central issues of modern cosmology. 
 
The most successful model in explaining such feature of our Universe and in fitting the available data (SnIa, BAO, CMB, Big Bang Nucleosynthesis (BBN) and Large Scale Structure (LSS)) is the  \textit{Standard Cosmological Model}, better known as $\Lambda$CDM \cite{Scott:2018adl, Akrami:2018vks}, in which the Universe is well described below the Planck scale by General Relativity (GR) and at scales of $100 \; \textrm{Mpc}$ and larger each component of the Universe can be described as a perfect fluid. The Universe is filled with baryonic matter, radiation, Cold Dark Matter (CDM) and the cosmological constant $\Lambda$, which is responsible for the late-time accelerated expansion \cite{Amendola:2015ksp, Calcagni:2017sdq}. 

Despite the success and the simplicity of the $\Lambda$CDM model, an explanation for the physical nature of CDM and $\Lambda$ is still missing and related to the latter, there are two big unsolved issues: it is still unclear why $\Lambda$ is so small compared to vacuum energy predicted by Quantum Field Theory (QFT) (fine-tuning problem) and why it becomes important only at very late times (coincidence problem) \cite{Weinberg:1988cp, Weinberg:2000yb}. 

More recently, as soon as we entered in the era of precision cosmology, tensions in the data started to appear. The most important example is the \textit{Hubble tension} \cite{Perivolaropoulos:2021jda}. On one hand, from the CMB anisotropy measurements from \textit{Planck} \cite{Aghanim:2018eyx}, for the value of the Hubble parameter $H_0$ we have $H_0 = (67.36 \pm 0.54) \; \textrm{km s}^{-1} \textrm{Mpc}^{-1}$, by assuming $\Lambda$CDM as the fiducial cosmological model.
On the other hand, the local measurements of the same parameter point towards greater values, among which from the \textit{Hubble} Space Telescope we have $H_0 = (74.03 \pm 1.42) \; \textrm{km s}^{-1} \textrm{Mpc}^{-1}$ \cite{Riess:2019cxk}, with a $4.4 \sigma$ discrepancy with the early-universe value above from \textit{Planck}.

Currently, this discrepancy does not seem to be due to systematic effects in either early-time or late-time measurements and may point to new physics beyond $\Lambda$CDM \cite{Bernal:2016gxb,Knox:2019rjx}.

Because of these issues, in the last decades, a great variety of models, which try to explain the late-time accelerated expansion of the Universe without resorting to the cosmological constant $\Lambda$, have been proposed \cite{Copeland:2006wr,Clifton:2011jh} which also try to address the Hubble tension problem \cite{DiValentino:2021izs}. For reviews on the experimental status of some of these theories beyond $\Lambda$CDM, see Refs. \cite{Joyce:2014kja,Huterer:2017buf}.

In most of the alternative models, the matter components filling the Universe are usually treated as perfect fluids. A slightly unconventional path to explain late-time accelerated expansion, which we consider in this paper is modifying the usual matter content by adding dissipative or viscous terms in the Energy-Momentum Tensor (EMT) describing perfect fluids. For recent reviews, see Refs. \cite{Brevik:2014cxa,Brevik:2017msy}.

Most of the best explored viscous models so far are those involving bulk viscosity where a term proportional to the Hubble function is added to one or more matter components of the Universe, giving rise to a modified effective pressure for them. These models may differ from one another in the assumptions made on the bulk viscosity coefficient and/or on the matter content of the Universe. For this reason, the effective pressure arising from viscosity can be important either in the early stages of the Universe or at the present time. 

In the first case, fall all those models in which the inflation is driven by viscosity \cite{Heller:1973Ap,Murphy:1973zz, Heller:1975Ap, Barrow:1986yf, Padmanabhan:1987dg, Maartens:1995wt, Maartens:1996dk, Chimento:2012xg, Myrzakulov:2014hva, Bamba:2015sxa, Szydlowski:2020awp}. A relevant feature of these models is that by admitting viscosity, one is able to avoid singularity in the Universe as well as to grow the Universe in size in oscillatory sequences. 

In the second case, the models that can describe the late-time accelerated expansion of the Universe are considered \cite{Meng:2005jy, Ren:2005nw,Brevik:2005bj, Szydlowski:2006ma, Avelino:2008rx, Avelino:2008ph, Li:2009mf, Velten:2011bg, Velten:2013rra, Brevik:2014eya,Barbosa:2015ndx,daSilva:2018ehn, Brevik:2019btm, MadrizAguilar:2020vsx}.  These models have revealed to be successful in reproducing the background evolution of the Universe at late time while attempting to solve the problems of $\Lambda$CDM, unify or replace the dark components of the Universe and be compatible with the large structure formation. It is still debated whether viscous terms in the pressure may or may not contribute to relieve the Hubble tension \cite{Anand:2017wsj,Yang:2019qza,Elizalde:2020mfs}.

Recently, with the purpose of describing the inflationary epoch of the Universe, a new framework, dubbed Ricci Cosmology, has been proposed \cite{Baier:2019ciw}. This framework involves second order corrections in gradients of the metric tensor $g_{\mu \nu}$ to the perfect fluid EMT, which arise in out-of-equilibrium relativistic fluid dynamics theory \cite{Romatschke:2017ejr}. 

In this paper we explore the viability of a simple solution in such a framework to describe at the background level the late-time accelerated expansion of the Universe, relieving at the same time the Hubble tension. The structure of the work is as follows. In Section \ref{sec:nequilfl}, we shortly review the relativistic fluid dynamics theory from which Ricci cosmology emerges. In Section \ref{sec:ricci}, we describe the features of our solution in the Ricci cosmology framework. 
In section \ref{sec:data}, we present statistical analysis and the data used to put constraints on the parameters of our model. In Section \ref{sec:results}, we show and discuss our results. In Section \ref{sec:conclusions}, we draw conclusions about the viability of the model in light of the cosmological data taken into account.

\section{\label{sec:nequilfl}Near-equilibrium Fluid Dynamics}
To understand the features of Ricci cosmology, it is useful to discuss some basics of fluid dynamics.

The first attempts to construct a consistent theory of dissipative fluids on a general curved spacetime are due to Eckart \cite{Eckart:1940te} and Landau and Lifshitz \cite{Landau1987Fluid}
in the first half of the twentieth century. After these first efforts which have been proven to be plagued with stability and causality problems, in the 1960s M\"{u}ller \cite{Muller:1967zza} and in the 1970s Israel and Stewart \cite{Israel:1976tn,Israel:1979wp} succeeded to solve these issues, as shown in \cite{Hiscock:1983zz}.

Since then, further developments occurred which culminated in the theory of relativistic fluid dynamics without conserved charges reviewed in \cite{Romatschke:2017ejr}, with an equivalent formulation for a fluid with a $U(1)$ symmetry based on an action principle, proposed by Kovtun et al. \cite{Kovtun:2018dvd}.
In the rest of the section, we will follow \cite{Romatschke:2017ejr}. 

\subsection{General construction}

A perfect fluid at the equilibrium in a curved spacetime is described by the Energy-Momentum Tensor (EMT)
\begin{equation}
    T^{\mu \nu}_{(0)} = \rho u^\mu u^\nu + P(\rho) h^{\mu \nu}
 \label{eq:0emt}
\end{equation}
where $\rho c^2$ is the energy density of the fluid as seen by a comoving observer with the fluid 4-velocity $u^\mu$ and $h^{\mu \nu} = g^{\mu \nu}+ u^\mu u^\nu/c^2$ is the 3-spatial metric of the hypersurface orthogonal to the fluid 4-velocity. The pressure $P(\rho)$ is the equilibrium pressure of the relativistic fluid,  whose form represents the equation of state (EoS) of the fluid. 
In general, a perfect fluid at the equilibrium has a barotropic EoS given by
\begin{equation}
    P(\rho) = w \rho c^2
\end{equation}
where $w$ is the constant EoS parameter depending on the nature of the fluid.

The energy density $\rho c^2$, the fluid 4-velocity $u^\mu$ and the metric $g_{\mu \nu}$ which fully characterise the description of the perfect fluid go under the name of hydrodynamic fields.

If the fluid is slightly out-of-equilibrium for the presence of dissipative effects or anisotropic non-stationary expansion, the perfect fluid description above is not enough to accurately characterize its dynamics, and gradients of such hydrodynamic fields are needed. 

In general, the EMT can be written as
\begin{equation}
    T^{\mu \nu}  = T_{(0)}^{\mu \nu} + T_{(1)}^{\mu \nu} + T_{(2)}^{\mu \nu} + \dots
\label{completeEMT}
\end{equation}
where the subscripts $(0), (1), (2), \dots$ indicate the number of gradients in each term of $T^{\mu \nu}$.

In order to obtain such higher order corrections to the perfect fluid EMT (\ref{eq:0emt}) scalars, vectors and tensors of such gradients are constructed and combined into tensors in such a way to split the contributions into a traceless part $\pi^{\mu \nu}$, referred to as the shear stress tensor
\begin{equation}
\pi^{\mu \nu} =  T^{<\mu \nu>}_{(1)} + T^{<\mu \nu>}_{(2)} + \dots
\label{constrel1}
\end{equation}
where $<\dots>$ indicates symmetrization over the indices and subtraction of the trace, and a trace part $\Pi h^{\mu \nu}$, with
\begin{equation}
\Pi = \frac{1}{d-1} \left( T^{\mu}_{(1) \mu} + T^{\mu}_{(2) \mu}  \right) + \dots
\label{constrel2}
\end{equation}
which is called the bulk stress.

The form of these corrections is further constrained by the conservation equation of motion for the full EMT
\begin{equation}
    \nabla_\mu T^{\mu \nu}=0
\end{equation}
and by the fact that the fluid energy flux is not modified by the corrections at any order, i.e.
\begin{equation}
    u_\mu T^{\mu \nu}_{(i)}=0,
\end{equation}
meaning that the energy density $\rho c^2$ seen by an observer comoving with the fluid is not modified by the non-equilibrium terms. This choice corresponds to the so-called \textit{Landau frame}.

Instead, from the corrections above, it is evident that the local pressure $P(\rho)$ departs from its equilibrium expression
\begin{equation}
p_{eff(i)} = P(\rho) + \pi^i_{i,LRF} + \Pi
\label{effpress}
\end{equation}
where in the second term there is no summation over $i$. 

When we neglect anisotropies, that is possible on cosmological scales which we are interested in here, the last equation simplifies to
\begin{equation}
p^{eff} = P(\rho)  + \Pi,
\label{eq:effpress1}
\end{equation}
which is an isotropic non-equilibrium pressure, due to internal friction in a fluid and can be responsible, in the cosmological context for the late-time accelerated expansion of the Universe.

At the first-order, the construction described above gives rise to the well-known expression
\begin{equation}
    T^{\mu \nu}_{(1)} = - \eta \sigma^{\mu \nu} - \zeta \Delta^{\mu \nu} \nabla_\lambda u^\lambda.
    \label{1stvisc}
\end{equation}
The term $\sigma^{\mu \nu} = 2 \nabla^{<\mu} u^{\nu>}$ is the well-known shear viscosity which becomes relevant in presence of anisotropies, and the second term is the bulk viscosity, which has been used in alternative cosmological models to Dark Energy and Inflation.

The shear viscosity coefficient $\eta$ and the bulk viscosity coefficient $\zeta$, are collectively called first order transport coefficients.  

By applying the same reasoning used for the first order correction $T^{\mu \nu}_{(1)}$, the second order correction to the perfect fluid EMT $T^{\mu \nu}_{(2)}$ can be obtained, by considering all the linearly independent scalars, vectors and rank-two tensors, containing exactly two gradients of the fundamental hydrodynamic fields and of the metric $g_{\mu \nu}$. The full details of the second order shear tensor $\pi^{\mu \nu}_{(2)}$ and of the second order bulk stress $\Pi_{(2)}$ can be found in \cite{Romatschke:2017ejr}.

Here, we are only interested in those terms in the second order bulk pressure $\Pi_{(2)}$ appearing  in the Ricci Cosmology framework proposed in \cite{Baier:2019ciw}, namely
\begin{equation}
    \Pi_{(2)} = \xi_5 R + \xi_6 u^\lambda u^\rho R_{\lambda \rho},
    \label{eq:2ndordconst}
\end{equation}
where $R_{\lambda \rho}$ is the Ricci tensor and $R$ is the Ricci scalar. The coefficients $\xi_5$ and $\xi_6$ are two of the second order transport coefficients, which parametrize the response of the non-equilibrium fluid to the change of the background. These terms are the only terms at the second order which are vanishing in Minkowski spacetime and non-zero in a curved spacetime. 

In the following, we will assume the second order transport coefficients and we will study their impact on the usual scaling of pressureless matter, radiation and cosmological constant with the redshift. 

\section{\label{sec:ricci}Ricci Cosmology}

\subsection{Ricci Cosmology model}
Let us assume that GR describes gravity in our Universe and it is well described at large scales by the FLRW metric
\begin{equation}
    ds^2 = -c^2 dt^2 +a^2(t) \delta_{ij} dx^i dx^j
\end{equation}
where $a(t)$ is the scale factor.

The Universe is assumed to undergo the usual sequence of radiation, matter and Dark Energy dominated epochs in which the contributions of the other matter components to the energy budget of the Universe is small and can be safely ignored.

The effective pressure of such matter components, due to the background expansion, gets modified by the terms in Eq.(\ref{eq:2ndordconst}) and it is given by
\begin{equation}
    p^{eff}_c = w_c \rho_c c^2+ \xi_{5c} R + \xi_{6c} u^\alpha u^\beta R_{\alpha \beta},
    \label{eq:effpress}
\end{equation}
where the subscript $c$ ranges on $m,r,\Lambda$ which stands for for matter, radiation and cosmological constant, respectively, and $\xi_{5c}$ and $\xi_{6c}$ are the second order transport coefficients assumed constant and different for every matter component.

The Ricci scalar $R$ and the projection of the Ricci tensor $R_{\alpha \beta}$ along the fluid 4-velocity $u^\alpha$ for the FLRW background are given by
\begin{equation}
\begin{split}
      R & = \frac{6}{c^2} \left( 2 H^2 + \dot{H} \right) \\
     u^\alpha u^\beta R_{\alpha \beta} & = R_{00} = -3 \left( H^2 + \dot{H} \right)
     \label{eq:ricciscal}
\end{split}
\end{equation}
where $H \equiv \frac{\dot{a}}{a}$ is the Hubble function.

The effective pressure in Eq.(\ref{eq:effpress}) can be seen to be analogous to the model of nonlinear bulk viscosity for the Dark Energy proposed in \cite{Ren:2005nw}. The two models coincide when the phenomenological parameter introduced in \cite{Ren:2005nw}, are $\zeta_0=0$, $\zeta_1 = -4 \xi_{5\Lambda}/c^2$ and $\zeta_2 = -4 \xi_{5 \Lambda}/c^2 + \xi_{6\Lambda}$, while the other coefficients are zero.

The conservation equation of the energy density $\rho_c c^2$  for such out-of-equilibrium cosmic fluid is given by
\begin{equation}
    \dot{\rho}_c + 3 H \left( \rho_c + \frac{p^{eff}_c}{c^2} \right) = 0,
    \label{eq:conseqnendens}
\end{equation}
with the energy density related to the Hubble function via the first Friedmann equation 
\begin{equation}
H^2 = \frac{8 \pi G}{3} \rho_c,
\label{eq:hubfuncsinglefl}
\end{equation}
in the epoch dominated by the matter component $c$. 

By inserting Eq.~(\ref{eq:effpress}) in Eq.~(\ref{eq:conseqnendens}), using the expressions in Eq.~(\ref{eq:ricciscal}), the equation for the energy density becomes
\begin{align}
    \dot{\rho}_c & + 3 H \Biggl[ \rho_c (1+w_c) \nonumber \\ &+ \frac{6 \xi_{5c}}{c^4} (2 H^2 + \dot{H}) - \frac{3 \xi_{6c}}{c^2} (H^2+\dot{H})\Biggr] =0.
     \label{eq:hubdiffeqn}
\end{align} 
By replacing cosmic time derivatives with redshift derivatives and using Eq.~(\ref{eq:hubfuncsinglefl}) to express the Hubble function in terms of the energy density, after rearranging the terms, the last equation becomes
\begin{align}
    -& (1+z) \frac{d \rho_c(z)}{dz} \left( 1+2 \hat{\xi}_{5c} - \hat{\xi}_{6c} \right) \nonumber \\ + & \left[ 3 (1 + w_c) + 8 \hat{\xi}_{5c} - 2 \hat{\xi}_{6c}\right] \rho_c(z) = 0
\label{eq:densitydiffeqn}
\end{align}
where the reduced second order transport coefficients
\begin{equation}
\label{eq:redcoeff}
    \hat{\xi}_{5c}\equiv \frac{12 \pi G \xi_{5c}}{c^4}, \quad \textrm{and} \quad \hat{\xi}_{6c} \equiv \frac{12 \pi G \xi_{6c}}{c^2},
\end{equation}
have been defined.
From the differential equation (\ref{eq:densitydiffeqn}), we find for the energy density the following expression 
\begin{equation}
    \rho_c (z) = \rho_{c0} (1+z)^{\frac{3 (1 + w_c) + 8 \hat{\xi}_{5c} - 2 \hat{\xi}_{6c}}{1+2 \hat{\xi}_{5c} - \hat{\xi}_{6c}}} \label{eq:density_split}
 \end{equation}
 Hence, the squared Hubble function for a flat Universe filled with matter, radiation and cosmological constant, all having a modified redshift scaling, reads
 \begin{equation}
    \frac{H^2 (z)}{H_0^2}= \Omega_m (1+z)^{3+\delta_m} + \Omega_r (1+z)^{4+\delta_r} + \Omega_\Lambda (1+z)^{\delta_\Lambda}
      \label{eq:hubfuncsquared}
\end{equation}
 where we have defined the dimensionless energy density parameter for the generic matter component $c$ as
 \begin{equation}
     \Omega_c \equiv \frac{\rho_{c0}}{\rho_{crit}}= \frac{8 \pi G}{3 H_0^2} \rho_{c0}
 \end{equation}
 with the relation $\Omega_\Lambda +\Omega_m + \Omega_r =1$ valid in a flat Universe.
 
Furthermore, the deviation parameters from usual scaling for matter, radiation and Cosmological Constant, i.e. $\delta_m$, $\delta_r$ and $\delta_\Lambda$, which appear in Eq.~(\ref{eq:hubfuncsquared}), can be easily derived from Eq.~(\ref{eq:density_split}) and are given by
\begin{equation}
     \delta_m =  \frac{2 \hat{\xi}_{5m} + \hat{\xi}_{6m}}{1+2 \hat{\xi}_{5m} - \hat{\xi}_{6m}},
    \label{eq:matdev}
\end{equation}
\begin{equation}
     \delta_r = \frac{2 \hat{\xi}_{6r}}{1+2 \hat{\xi}_{5r} - \hat{\xi}_{6r}},
     \label{eq:raddev}
\end{equation}
and
\begin{equation}
   \delta_\Lambda =  \frac{8 \hat{\xi}_{5 \Lambda} - 2 \hat{\xi}_{6\Lambda}}{1+2 \hat{\xi}_{5\Lambda} - \hat{\xi}_{6\Lambda}},
   \label{eq:cosmdev}
\end{equation}
respectively.

\subsection{Thermodynamical priors}

Before entering the observational bounds on the parameters of our models, we impose some physical bounds on them by applying the Second Law of Thermodynamics which says that the entropy of a system $S$ never decreases \cite{Barrow:1988yc}. These bounds can then be treated as priors for further statistical considerations. 

For a fluid with energy $E= \rho c^2  V$ in a volume $V$, with temperature $T$ and pressure $p$, from the First Law of Thermodynamics 
\begin{equation}
    dE= T dS -p dV
\end{equation}
the conservation equation for the energy density of such a fluid can be derived
\begin{equation}
    \dot{\rho} + 3 H \left( \rho + \frac{p}{c^2} \right) - \frac{T}{V} \frac{\dot{S}}{c^2} = 0
\end{equation}
By comparing the last equation with Eqs. (\ref{eq:effpress}) and (\ref{eq:conseqnendens}), we find a the following differential equation for the entropy $S$
\begin{equation}
    \frac{T}{V} \dot{S} = - 3 H  \left( \xi_{5} R +\xi_{6} u^\alpha u^\beta R_{\alpha \beta} \right) 
\end{equation}
From the thermodynamic relation for enthalpy $ H \equiv E + p V = T S$, the entropy can be expressed as
\begin{equation}
    S = \left( \rho c^2 +p \right) \frac{V}{T}
\end{equation}
By dividing the last two equations, we arrive at
\begin{equation}
    \frac{\dot{S}}{S} = - \frac{3H}{\rho c^2 +p} \left( \xi_{5} R +\xi_{6} u^\alpha u^\beta R_{\alpha \beta} \right) 
\end{equation}
This must be valid for each matter component filling the Universe.

For a matter component with energy density $\rho_c c^2$, barotropic EoS parameter $w_c$ and the constant reduced second order transport coefficients $\hat{\xi}_{5c}$ and $\hat{\xi}_{6c}$, the entropy $S$ in terms of the scale factor, after a simple calculation, reads
\begin{equation}
    S(a) = S_0 
    a^{- \frac{1}{1+w_c} \left[ 8 \hat{\xi}_{5c}- 2 \hat{\xi}_{6c} {-} \delta_c (2 \hat{\xi}_{5c}- \hat{\xi}_{6c} ) \right] + 3 (2 \hat{\xi}_{5c}- \hat{\xi}_{6c})}.
    \label{eq:entropyeqn}
\end{equation}
As we have already mentioned, from the Second Law of Thermodynamics for an isolated system, for each epoch of the Universe, there must be an increase of entropy
\begin{equation}
\Delta S \geq 0 .
\end{equation}

Thus, by specializing Eq.(\ref{eq:entropyeqn}) to each matter components of the Universe and using the expressions for $\delta_m$, $\delta_r$ and $\delta_\Lambda$ found above, we have direct or indirect constraints on the deviation parameters. 

For the cosmological constant, the entropy is given by 
    \begin{equation}
         S(a) = S_0 a^{ 3 (2 \hat{\xi}_{5 \Lambda}- \hat{\xi}_{6\Lambda})} ,
    \end{equation}
    which increases for $2\hat{\xi}_{5 \Lambda}- \hat{\xi}_{6 \Lambda} \geq 0$.
    
For pressureless matter, the entropy reads
\begin{equation}
    S(a) = S_0 a^{- \delta_m} ,
\end{equation}
which increases for $\delta_m \leq 0$.

Finally, for radiation, it holds
\begin{equation}
    S(a) = S_0 a^{- \frac{3}{4} \delta_r} ,
\end{equation}
which increases for $\delta_r \leq 0$.

In the next section, we describe the statistical analysis and the data we use to put bounds on the parameters of our model. 

\section{\label{sec:data} Statistical Analysis \& Data}

In order to assess the viability of the model described
in the previous section and determine the relative importance of the two contributions to the out-of-equilibrium, we test against data four special cases of the model described above, with four different ans\"{a}tze on the transport coefficients.

In the Table \ref{tab:tablemodels}, we report the assumptions made on the constant second order transport coefficients with physical priors on the deviation parameters derived from the physical requirement of the increase of entropy, we use in our statistical analysis, together with the usual priors on the cosmological parameters in common with $\Lambda$CDM ($0< \Omega_b< \Omega_m<1$, $0<h<1$).

{\renewcommand{\arraystretch}{1.8}
\begin{table*}
\begin{ruledtabular}
\begin{tabular}{ccc}
 & Assumptions on $\hat{\xi}_5$ and $\hat{\xi}_6$ & Priors on $\delta_m$, $\delta_r$ and $\delta_\Lambda$ \\
\hline
Ansatz 1 & $\hat{\xi}_{5r} = \hat{\xi}_{6r} = 0$ and $\hat{\xi}_{5m} = \hat{\xi}_{5\Lambda}\equiv \hat{\xi}_{50}$,  $\hat{\xi}_{6m} = \hat{\xi}_{6\Lambda}\equiv \hat{\xi}_{60}$ & $\delta_r=0$, $\delta_m \leq 0$ and $\delta_m \leq \delta_\Lambda < 3 + \delta_m$ \\
\hline
Ansatz 2 & $\hat{\xi}_{5r} =\hat{\xi}_{5m} = \hat{\xi}_{5\Lambda}\equiv \hat{\xi}_{50}$ and  $\hat{\xi}_{6r} =  \hat{\xi}_{6m} = \hat{\xi}_{6\Lambda}\equiv \hat{\xi}_{60}$ & $\delta_\Lambda = 4 \delta_m - 3 \delta_r$, $\delta_m \leq 0$ and $-1 +\delta_m<\delta_r \leq \delta_m$ \\
\hline
Ansatz 3  & $\hat{\xi}_{5r} = \hat{\xi}_{6r} = 0$ and $\hat{\xi}_{5m} = \hat{\xi}_{6m}$, $\hat{\xi}_{5\Lambda}= \hat{\xi}_{6\Lambda}$ & $\delta_r=0$, $\delta_m \leq 0$ and $\delta_\Lambda \geq 0$ \\
\hline
Ansatz 4 & $\hat{\xi}_{5r} = \hat{\xi}_{6r} = 0$ and $\hat{\xi}_{5m} = \hat{\xi}_{5\Lambda}\equiv \hat{\xi}_{50}$,  $\hat{\xi}_{6m} = \hat{\xi}_{6\Lambda}\equiv \hat{\xi}_{60}$ & $\delta_r \leq 0$, $\delta_m \leq 0$ and $\delta_\Lambda \geq 0$  \\
\end{tabular}
\end{ruledtabular}
\caption{\label{tab:tablemodels} The table reports the assumptions on the reduced second order transport coefficients and the priors on the deviation parameters $\delta_m, \delta_r$ and $\delta_\Lambda$ appearing in the Hubble function for each ansatz on the transport coefficients imposed on our model, derived in the appendix \ref{sec:appendix}.}
\end{table*}
}

In the first two ans\"{a}tze, we have that the Ricci scalar and the time-time component of the Ricci  tensor in the effective pressure have the same effect in the change of the scaling of the matter components filling
the Universe. The difference between the two cases is in the way radiation is treated: in ansatz 1, radiation is assumed to be unaffected by the modifying pressure terms, while in ansatz 2, radiation fluid deviates from conformality due to the pressure terms.

In the last two ans\"{a}tze, instead, each fluid differs from the other for the transport coefficients which characterise its response to both Ricci scalar and time-time component of the Ricci tensor. Analogously to the previous ans\"{a}tze, we consider two different behaviours for radiation.

In the fits of the special cases of our model, we combine the following data sets we have on the background evolution at large scales of the Universe: Type Ia Supernovae (SNeIa) from the Pantheon sample \cite{Scolnic:2017caz}, the Mayflower sample of Gamma Ray Bursts (GRBs) \cite{Liu:2014vda}, the Early-Type Galaxies (ETG) used as Cosmic Chronometers (CC) \cite{Moresco:2012by,Moresco:2015cya,Gomez-Valent:2019lny}, the data on the Hubble parameter $H_0$ from the H0LiCOW collaboration \cite{Suyu:2016qxx,Wong:2019kwg}, Baryon Acoustic Oscillations (BAO) from different surveys \cite{Hinton:2016atz,Ata:2017dya, Alam:2016hwk, Agathe:2019vsu, Nadathur:2019mct}, and the last \textit{Planck} release for the Cosmic Microwave Background (CMB) \cite{Zhai:2018vmm}.

In order to evaluate the performance of our model in relieving the Hubble tension, we consider two different combinations of our data sets: the late time data set including only late time measurements of cosmological observables (SNeIa, CC,
H0LiCOW, GRBs and BAO from WiggleZ), and the full data set which combines these observations with early time observations (CMB and BAO data from SDSS).

To find the best fit parameters of our model for each ansatz, we minimise the total  $\chi^2$ given by the sum of all the $\chi^2$ associated with the data from the probes listed above
\begin{equation}
\chi^2 = \chi^2_{SN}+ \chi_G^2 + \chi_H^2 + \chi^2_{HCOW} + \chi^2_{BAO} + \chi^2_{CMB}
\end{equation}
by using our own implementation of a Monte Carlo Markov Chain (MCMC).

Then, to see under which assumptions on the transport coefficients our model can fit the data better than $\Lambda$CDM, we have fitted $\Lambda$CDM to the same data sets and we have compared each of the ans\"{a}tze on our model to it by means of the Bayes factor. 

Given a generic model $M_i$, with $\pi(\boldsymbol{\theta}_i|M_i)$, the prior probability of its set of parameters $\boldsymbol{\theta}_i$, and its likelihood function $\mathcal{L}_i (D | \boldsymbol{\theta}_i, M_i) \propto e^{-\chi^2/2}$, the Bayesian evidence $\mathcal{E}_i$ is defined as the probability of the data $D$ given the model $M_i$ with a set of parameters $\boldsymbol{\theta}_i$
\begin{equation}
\mathcal{E}_i = \int d \theta_i \mathcal{L}_i (D | \boldsymbol{\theta}_i, M_i) \pi (\boldsymbol{\theta}_i|M_i),
\end{equation}

Then, to compare the fit of this model to that of another model $M_j$, tested against the same set of data, in general depending on a different set of parameters $\boldsymbol{\theta}_j$, we compute the Bayes factor defined as
\begin{equation}
 \mathcal{B}^i_j = \frac{\mathcal{E}_i}{\mathcal{E}_j},
\end{equation}
where $\mathcal{E}_j$ is the Bayesian evidence of the model $M_j$.
The last model $M_j$, in our case, is $\Lambda$CDM while the model $M_i$ is one of the four special cases of our model.

Then, we can evaluate the performance of our model with respect to $\Lambda$CDM in fitting the data for each ansatz considered, by using the Jeffreys’ Scale \cite{Trotta:2005ar}: if $\ln \mathcal{B}^i_j < 1$, the evidence in favour of model $M_i$ is not significant; if  $1<\ln \mathcal{B}^i_j <2.5$, it is substantial; if  $2.5<\ln \mathcal{B}^i_j <5$, it is strong; if  $\ln \mathcal{B}^i_j >5$, it is decisive, while negative values of $\ln \mathcal{B}^i_j$ can be instead interpreted as evidence against model $M_i$ and thus, in favour of model $M_j$.

\subsection{Type Ia Supernovae}

The Pantheon catalogue contains 1048 Type Ia Supernovae (SnIa), used as \textit{standard candles}, in the redshift interval $0.01 <z<2.26$ \cite{Scolnic:2017caz}. 

The relevant observable for this data set is the distance modulus $\mu_{SN}$, defined as 
\begin{equation}
\mu_{SN} = m_{SN} - M_{SN}
\end{equation}
where $m_{SN}$ is the apparent magnitude and $M_{SN}$ is the absolute magnitude of the SnIa. 

Given $\boldsymbol{\theta}$, the vector of the cosmological parameters of the model under consideration, the distance modulus $\mu_{SN}$ is related to the luminosity distance $d_L$, by the formula
\begin{equation}
    \mu_{SN}(z, \boldsymbol{\theta}) = 5 \log d_L(z, \boldsymbol{\theta}) +\mu_0.
    \label{eq:distmodulus}
\end{equation}
where $\mu_0$ is an offset parameter which we marginalize over following Ref. \cite{Conley:2011ku}, and the luminosity distance $d_L$ is given by
\begin{equation}
    d_L(z, \boldsymbol{\theta}) = \frac{c}{H_0} (1+z) \int_0^z \frac{dz'}{E(z',\boldsymbol{\theta})},
\end{equation} 
where $E(z, \boldsymbol{\theta}) = H(z, \boldsymbol{\theta})/H_0$. 

The total $\chi^2_{SN}$ associated to this sample is
\begin{equation}
    \chi^2_{SN} = \Delta \boldsymbol{\mu}_{SN} \boldsymbol{C}^{-1}_{SN} \Delta \boldsymbol{\mu}_{SN} 
\end{equation}
where $\boldsymbol{C}_{SN}$ is the the covariance matrix and $\Delta \boldsymbol{\mu}_{SN} = \boldsymbol{\mu}_{SN} - \boldsymbol{\mu}_{SN}^{obs}$ is the difference between the theoretical value and the observed one of the distance modulus of the supernovae in the sample.

\subsection{Gamma Ray Bursts}
We consider 79 Gamma Ray Bursts (GRBs) from the Mayflower sample in the redshift range $1.44 <z<8.1$ \cite{Liu:2014vda}. 
Analogously to the case of the Type Ia Supernovae, the observable for the GRBs is the distance modulus $\mu_{GRB}$, defined as in Eq.(\ref{eq:distmodulus}), with the corresponding offset parameter $\mu_0$ marginalized over as for the Type Ia Supernovae.

The total $\chi^2_{GRB}$ for this sample is given by
\begin{equation}
    \chi^2_{GRB} = \Delta \boldsymbol{\mu}_{GRB} \boldsymbol{C}^{-1}_{GRB} \Delta \boldsymbol{\mu}_{GRB} 
\end{equation}
where $\Delta \mu_{GRB} = \mu_{GRB} - \mu_{GRB}^{obs}$.

\subsection{Cosmic Chronometers}
Passively-evolving Early-Type Galaxies (ETGs), i.e. galaxies with low star formation rate and old stellar populations, can be used as \textit{standard clocks} or Cosmic Chronometers (CC). \\ Differently from other probes, they give us direct information about the Hubble function over redshift ranges, from the formula
\begin{equation}
    H(z) = - (1+z)^{-1} \frac{dz}{dt}
\end{equation}
where $\frac{dz}{dt}$ is inferred from observations. \\
The sample used in our data analysis covers the range $0<z<1.97$ \cite{Moresco:2012by,Moresco:2015cya,Gomez-Valent:2019lny}. \\
The $\chi^2_H$ reads as
\begin{equation}
    \chi^2_H = \sum_{i=1}^{29} \frac{(H(z_i, \boldsymbol{\theta}) - H_{obs}(z_i))^2}{\sigma^2_H(z_i)}
\end{equation}
where $\sigma_H(z_i)$ are the statistical errors on the measured values of the Hubble function $H_{obs}(z_i)$.

\subsection{H0LiCOW}
The H0LiCOW collaboration \cite{Suyu:2016qxx} uses 6 strong gravitationally lensed quasars with multiple images to put constraints on the value of the Hubble parameter $H_0$ \cite{Wong:2019kwg}. \\
In order to constrain the parameters of our model, we use the so-called
\textit{time-delay distance}, which is defined by
\begin{equation}
    D_{\Delta t} \equiv (1+z_L) \frac{D_L D_S}{D_{LS}}
\end{equation}
where $z_L$ is the lens redshift, and $D_S$, $D_L$ and $D_{LS}$ represent the angular diameter distances from the source to the observer, from the lens to the observer, and between source and lens, respectively, with the angular diameter distance given by
\begin{equation}
    D_A(z,\boldsymbol{\theta}) = \frac{c}{H_0}\frac{1}{1+z} \int_0^z \frac{c dz'}{E(z',\boldsymbol{\theta})}.
    \label{eq:angdiamdist}
\end{equation}
The $\chi^2$ for the H0LiCOW data
\begin{equation}
    \chi^2_{HCOW} = \displaystyle{\sum_{i=1}^6} \frac{(D_{\Delta t,i} (\boldsymbol{\theta}) - D^{obs}_{\Delta t,i})^2}{\sigma^2_{D_{\Delta t,i}}}
\end{equation}
where $\sigma_{D_{\Delta t,i}}$ are the statistical errors on the measured time-delay distances $D^{obs}_{\Delta t,i}$.

\subsection{Baryon Acoustic Oscillations}
In the following, we report the five BAO data sets used in our analysis.

From the WiggleZ Dark Energy Survey \cite{Blake:2012pj} the physical observables taken into account are the acoustic parameter
\begin{equation}
    A(z, \boldsymbol{\theta}) = 100 \sqrt{\omega_m} \frac{D_V(z, \boldsymbol{\theta})}{c z}
\end{equation}
with $\omega_m = \Omega_m h^2$, and the Alcock-Paczynski distortion parameter
\begin{equation}
    F(z, \boldsymbol{\theta}) = (1+z) \frac{D_A(z, \boldsymbol{\theta}) H(z, \boldsymbol{\theta}) }{c}
\end{equation}
where $D_A$ is the angular diameter distance defined in Eq.(\ref{eq:angdiamdist}) and $D_V$ is the volume distance given by
\begin{equation}
    D_V(z, \boldsymbol{\theta}) = \left[ (1+z)^2 D_A^2(z, \boldsymbol{\theta}) \frac{cz}{H(z, \boldsymbol{\theta})}\right]^{1/3}.
\end{equation}
From the SDSS-III BOSS DR12 \cite{Alam:2016hwk}, the quantities 
\begin{equation}
    D_M(z, \boldsymbol{\theta}) \frac{r^{fid}_s(z_d)}{r_s(z_d,  \boldsymbol{\theta})}, \quad H(z) \frac{r_s(z_d,  \boldsymbol{\theta})}{r^{fid}_s(z_d)}
\label{sdssiiiboss}
\end{equation}
are considered, where $D_M$ denotes the comoving distance
\begin{equation}
    D_M(z, \boldsymbol{\theta}) = \frac{c}{H_0} \int^z_0 \frac{1}{E(z', \boldsymbol{\theta})} dz'
\end{equation}
and $r_s$ is the sound horizon
\begin{equation}
    r_s(z, \boldsymbol{\theta}) = \int_z^\infty \frac{c_s(z')}{H(z', \boldsymbol{\theta})} dz',
    \label{eq:soundhorizon}
\end{equation}
which is evaluated at the dragging redshift $z_d$ in Eq.\eqref{sdssiiiboss}. 

In the same expression, $r^{fid}_s (z_d)$ is the sound horizon evaluated at the dragging redshift $z_d$, considering a  fiducial cosmological model.

 The dragging redshift $z_d$ can be estimated numerically as \cite{Eisenstein:1997ik}
\begin{equation}
    z_d= 1291 \frac{\omega_m^{0.251}}{1+0.659 \; \omega_m^{0.828}}[1+b_1 \; \omega_b^{b_2}]
\end{equation}
with $\omega_b= \Omega_b h^2$ and the coefficients $b_1$ and $b_2$ given by
\begin{equation}
    b_1=0.313 \; \omega_m^{-0.419} [1+0.607 \; \omega_m^{0.6748}]
\end{equation}
\begin{equation}
    b_2 = 0.238 \; \omega_m^{0.223},
\end{equation}
respectively.

The sound speed for coupled photons and baryons, appearing in Eq.(\ref{eq:soundhorizon}), in $\Lambda$CDM is given by
\begin{equation}
    c_s(z) = \frac{c}{\sqrt{3(1+\bar{R}_b (1+z)^{-1}})},
\end{equation}
where $\bar{R}_b$ is baryon-to-photon density ratio parameter, given by 
\begin{equation}
    \bar{R}_b = 31500 \Omega_b h^2 (T_{CMB} /2.7)^{-4}
\end{equation}
with the CMB temperature $T_{CMB} = 2.726 \; \mathrm{K}$, while for our model, it must be modified as
\begin{equation}
  c_s(z) = \frac{c}{\sqrt{3\left( 1+ \bar{R}_b (1+z)^{-1-\delta_r+\delta_m} \right)}}
\end{equation}
with  
\begin{equation}
    \bar{R}_b = 31500 \Omega_b h^2 \left(\frac{1+\frac{\delta_m}{3}}{1+\frac{\delta_r}{4}} \right) \left( \frac{T_{CMB}}{2.7} \right)^{-4},
\end{equation}
due to the modified scaling of matter and radiation parametrized by $\delta_m$ and $\delta_r$, respectively.

The other measurements for BAO used in the data analysis are the following:
\begin{itemize}
\item from the combination of void-galaxy cross-correlation with BAO and galaxy RSD in the CMASS
galaxy catalog of the BOSS DR12 \cite{Nadathur:2019mct}
\begin{equation}
    \frac{D_A(z=0.57)}{r_s(z_d)}= 9.383 \pm 0.077,
\end{equation}
\begin{equation}
    H(z=0.57) r_s(z_d) = (14.05 \pm 0.14) 10^3 \; \textrm{km/s}
\end{equation}
\item from eBOSS DR14, a spherically-averaged BAO distance \cite{Ata:2017dya}
\begin{equation}
    D_V(z=1.52)= 3843 \pm 147 \frac{r_s(z_d)}{r^{fid}_s(z_d)} \; \textrm{Mpc}
\end{equation}
\item from eBOSS DR14, by combining the quasar Lyman-$\alpha$ autocorrelation function with the quasar Lyman-$\alpha$
cross-correlation measurement \cite{Blomqvist:2019rah,Agathe:2019vsu}
\begin{equation}
    \frac{D_A(z= 2.34)}{r_s(z_d)} = 36.98^{+1.26}_{-1.18}
\end{equation}
\begin{equation}
    \frac{c}{H(z=2.34) r_s(z_d)} = 9.00 \pm 0.22
\end{equation}
\end{itemize}
For each BAO probe, the $\chi^2$ is given by 
\begin{equation}
    \chi^2_{BAO} = \Delta \boldsymbol{X}^{BAO} \boldsymbol{C}^{-1}_{BAO} \Delta \boldsymbol{X}^{BAO}
\end{equation}
where $\Delta \boldsymbol{X}^{BAO} = \boldsymbol{X}^{BAO} - \boldsymbol{X}^{BAO}_{obs}$ is the difference between the predicted and observed values for the observables of each probe.

\subsection{Cosmic Microwave Background}
The last data set we use is given by the Cosmic Microwave Background (CMB).

The CMB data we include in our data set are the shift parameters \cite{Wang:2007mza}
from the last \textit{Planck} data release \cite{Zhai:2018vmm}: the physical baryon density parameter $\omega_b$, the angular scale of the sound horizon at recombination
\begin{equation}
    l_a (\boldsymbol{\theta}) \equiv \pi \frac{D_M(z_*, \boldsymbol{\theta})}{r_s (z_*, \boldsymbol{\theta})},
\end{equation}
and the scaled distance to recombination
\begin{equation}
    R(\boldsymbol{\theta}) \equiv \sqrt{\Omega_m H_0^2} \frac{D_M(z_*,\boldsymbol{\theta})}{c}
\end{equation}
with the sound horizon $r_s$ and $D_M$ evaluated at the recombination redshift $z_*$, which is given by the fitting formula \cite{Hu:1995en}
\begin{equation}
    z_* = 1048[1+0.00124 \; \omega_b^{-0.738}] (1+g_1 \; \omega_m^{g_2})
\end{equation}     
with the factors $g_1$ and $g_2$ given by
\begin{equation}
    g_1 = \frac{0.0783 \; \omega_b^{-0.238}}{1+39.5 \; \omega_b^{-0.763}}
\end{equation}
\begin{equation}
    g_2 = \frac{0.560}{1+ 21.1 \; \omega_b^{1.81}},
\end{equation}
respectively.

Therefore, the $\chi^2$ for the CMB data is 
\begin{equation}
    \chi^2_{CMB} = \Delta \boldsymbol{X}^{CMB} \boldsymbol{C}_{CMB}^{-1}  \Delta \boldsymbol{X}^{CMB}
\end{equation}
where $\Delta \boldsymbol{X}^{CMB} = \boldsymbol{X}^{CMB} - \boldsymbol{X}_{obs}^{CMB}$ is the difference between the predicted and observed values for the quantities in the vector $\boldsymbol{X}=\{ \omega_b, l_a,R \}$.

\section{\label{sec:results} Results {\&} Discussion} 

{\renewcommand{\arraystretch}{1.7}
\begin{table*}
\begin{ruledtabular}
\begin{tabular}{ccccc}
& $\Lambda\text{CDM}$ & $\Lambda\text{CDM}$ & $\text{Ansatz 1}$ & $\text{Ansatz 1}$ \\
& full & late & full & late \\
\hline
$h$ & $0.673^{+0.003}_{-0.003}$ & $0.713^{+0.013}_{-0.012}$ & $0.668^{+0.004}_{-0.005}$ & $0.713^{+0.014}_{-0.014}$ \\
$\Omega_m$ & $0.319^{+0.005}_{-0.005}$ &
$0.292^{+0.017}_{-0.016}$ & $0.319^{+0.005}_{-0.005}$ & $0.307^{+0.023}_{-0.022}$ \\
$\delta_m$ & & & $>-0.001$ & $>-0.241$ \\
$\delta_\Lambda$ & & & $<0.003$ & $0.141^{0.192}_{-0.153}$ \\
\hline
$\Omega_\Lambda$ & $0.681^{+0.005}_{-0.005}$ &
$0.708^{+0.016}_{-0.017}$ & $0.681^{+0.005}_{-0.005}$ &
$0.693^{+0.022}_{-0.023}$ \\
$\xi_{50} \textrm{(N)}$ & & &  $0.96^{9.57}_{-8.69} \times 10^{38}$ & $-7.08^{6.37}_{-6.89} \times 10^{40}$ \\
$\xi_{60} \textrm{(kg/m)}$ & & & $> -4.66 \times 10^{22}$ &  $>-7.90 \times 10^{24}$ \\
\hline
$\ln \mathcal{B}^i_j$ & $0$ & $0$ & $-1.56^{+0.04}_{-0.04}$ & $-0.59^{+0.03}_{-0.03}$ \\
\end{tabular}
\end{ruledtabular}
\caption{\label{tab:table1} In the table, we report the constraints on our model parameters for ansatz $1$ and $\Lambda$CDM, both tested against the same combination of data sets. From the reported Bayes factors, our model with ansatz $1$ is disfavoured with respect to $\Lambda$CDM for both late and full data sets.}
\end{table*}}

{\renewcommand{\arraystretch}{1.7}
\begin{table*}
\begin{ruledtabular}
\begin{tabular}{ccccc}
& $\Lambda\text{CDM}$ &  $\Lambda\text{CDM}$ & $\text{Ansatz 2}$ & $\text{Ansatz 2}$\\
& full & late & full & late \\
\hline
$h$ & $0.673^{+0.003}_{-0.003}$ &  $0.713^{+0.013}_{-0.012}$ & $0.667^{+0.005}_{-0.005}$ & $0.713^{+0.013}_{-0.013}$ \\
$\Omega_m$ & $0.319^{+0.005}_{-0.005}$ & $0.292^{+0.017}_{-0.016}$ & $0.319^{+0.005}_{-0.005}$ & $0.306^{+0.023}_{-0.022}$ \\
$\delta_m$ & & & $> -0.0007$ & $>-0.244$ \\
$\delta_r$ & & & $>-0.003$ & $>-0.388$ \\
\hline
$\delta_\Lambda$ & & & $<0.007$ & $0.151^{+0.189}_{-0.160}$ \\
$\Omega_\Lambda$ & $0.681^{+0.005}_{-0.005}$ & $0.708^{+0.016}_{-0.017}$ & $0.681^{+0.005}_{-0.005}$ & $0.694^{+0.022}_{-0.023}$ \\
$\xi_{50} \textrm{(N)}$ & & & $0.79^{1.59}_{-0.93} \times 10^{39}$ & $-6.97_{-6.85}^{6.27} \times 10^{40}$\\
$\xi_{60} \textrm{(kg/m)}$ & & & $>-5.39 \times 10^{22}$ & $>-8.13 \times 10^{24}$\\
\hline
$\ln \mathcal{B}^i_j$ & $0$ & $0$ & $-1.56^{+0.04}_{-0.03}$ & $-0.59^{+0.02}_{-0.03}$\\
\end{tabular}
\end{ruledtabular}
\caption{\label{tab:table2}In the table, we report the constraints on our model parameters for ansatz $2$ and $\Lambda$CDM, both tested against the same combination of data sets. From the reported Bayes factors, our model with ansatz $2$ is disfavoured with respect to $\Lambda$CDM for both late and full data sets.}
\end{table*}}

{\renewcommand{\arraystretch}{1.7}
\begin{table*}
\begin{ruledtabular}
\begin{tabular}{ccccc}
& $\Lambda\text{CDM}$ & $\Lambda\text{CDM}$ & $\text{Ansatz 3}$ & $\text{Ansatz 3}$ \\
& full & late & full & late \\
\hline
$h$ & $0.673^{+0.003}_{-0.003}$ & $0.713^{+0.013}_{-0.012}$ & $0.667^{+0.004}_{-0.007}$ & $0.711^{+0.013}_{-0.013}$\\
$\Omega_m$ & $0.319^{+0.005}_{-0.005}$ & $0.292^{+0.017}_{-0.016}$ & $0.319^{+0.005}_{-0.005}$ & $0.304^{+0.023}_{-0.021}$\\
$\delta_m$ & & & $>-0.001$ & $>-0.256$ \\
$\delta_\Lambda$ & & & $< 0.004$ & $<0.277$ \\
\hline
$\Omega_\Lambda$ & $0.681^{+0.005}_{-0.005}$ & $0.708^{+0.016}_{-0.017}$ & $0.681^{+0.005}_{-0.005}$ & $0.696^{+0.021}_{-0.023}$\\
$\xi_{5 m} \textrm{(N)}$ & & & $> -1.03 \times 10^{39}$ & $> -2.53 \times 10^{41}$ \\
$\xi_{5\Lambda} \textrm{(N)}$ & & & $<2.35 \times 10^{39}$ & $<1.55 \times 10^{41}$ \\
$\xi_{6m} \textrm{(kg/m)}$ & & & $> -1.14 \times 10^{22}$ & $>-2.20 \times 10^{24}$ \\
$\xi_{6\Lambda} \textrm{(kg/m)}$ & & & $< 2.62 \times 10^{22}$ & $<1.72 \times 10^{24}$\\
\hline
$\ln \mathcal{B}^i_j$ & $0$ & $0$ & $-1.53^{+0.03}_{-0.04}$ & $-0.68^{+0.02}_{-0.03}$ \\
\end{tabular}
\end{ruledtabular}
\caption{\label{tab:table3}In the table, we report the constraints on our model parameters for ansatz $3$ and $\Lambda$CDM, both tested against the same combination of data sets. From the reported Bayes factors, our model with ansatz $3$ is disfavoured with respect to $\Lambda$CDM for both late and full data sets.}
\end{table*}}

{\renewcommand{\arraystretch}{1.7}
\begin{table*}
\begin{ruledtabular}
\begin{tabular}{ccccc}
& $\Lambda\text{CDM}$  & $\Lambda\text{CDM}$ & $\text{Ansatz 4}$ & $\text{Ansatz 4}$ \\
& full & late & full & late \\
\hline
$h$ & $0.673^{+0.003}_{-0.003}$ & $0.713^{+0.013}_{-0.012}$ & $0.701^{+0.012}_{-0.011}$ & $0.712^{+0.013}_{-0.013}$ \\
$\Omega_m$ & $0.319^{+0.005}_{-0.005}$ & $0.292^{+0.017}_{-0.016}$ & $0.322^{+0.005}_{-0.005}$ & $0.303^{+0.023}_{-0.020}$ \\
$\delta_m$ & & & $>-0.002$ & $> - 0.245$ \\
$\delta_r$ & & & $>-0.001$ & $> -199$ \\
$\delta_\Lambda$ & & & $<0.091$ & $< 0.269$  \\
\hline
$\Omega_\Lambda$ & $0.681^{+0.005}_{-0.005}$ & $0.708^{+0.016}_{-0.017}$ & $0.678^{+0.005}_{-0.005}$ & $0.697^{+0.020}_{-0.023}$ \\
$\xi_{5 m} \textrm{(N)}$ & & & $>-1.86 \times 10^{39}$ & $>-2.42 \times 10^{41}$ \\
$\xi_{5 r} \textrm{(N)}$ & & & $>-1.76 \times 10^{39}$ & $>-3.18 \times 10^{42}$ \\
$\xi_{5\Lambda} \textrm{(N)}$ & & & $<4.95 \times 10^{40}$ & $<1.50 \times 10^{41}$\\
$\xi_{6 m} \textrm{(kg/m)}$ & & & $>-2.08 \times 10^{22}$ & $>-2.72 \times 10^{24}$ \\
$\xi_{6r} \textrm{(kg/m)}$ & & & $>-1.96 \times 10^{22}$ & $>-3.54 \times 10^{25}$\\
$\xi_{6\Lambda} \textrm{(kg/m)}$ & & & $<5.53 \times 10^{23}$ & $< 1.67 \times 10^{24}$ \\
\hline
$\ln \mathcal{B}^i_j$ & 0 & 0 & $-2.01^{+0.04}_{-0.03}$ & $-0.63^{+0.04}_{-0.03}$ \\
\end{tabular}
\end{ruledtabular}
\caption{\label{tab:table4}In the table, we report the constraints on our model parameters for ansatz $4$ and $\Lambda$CDM, both tested against the same combination of data sets. From the reported Bayes factors, our model with ansatz $4$ is disfavoured with respect to $\Lambda$CDM for both late and full data sets.}
\end{table*}}

The results of the fits of our model for the four ans\"{a}tze are reported in the Tables \ref{tab:table1}, \ref{tab:table2}, \ref{tab:table3} and \ref{tab:table4}.

At $1 \sigma$ level, we note that the values of the cosmological parameters are indistinguishable from those corresponding to the standard $\Lambda$CDM and so we have not found a relief of the Hubble tension with any of our ans\"{a}tze.

From the fits with late time data sets, we have found milder upper or lower bounds on the deviation parameters of our model, with respect to the bounds obtained by using the full data set comprising late and early-time cosmological data which show more constraining power. The only exception is represented by $\delta_\Lambda$ for the ans\"{a}tze $1$ and $2$ which is compatible with zero in the fits with only late time data sets.

Regarding the second order transport coefficients, the fits of the model with the first two ans\"{a}tze imply that $\xi_{50}$ is compatible with zero, while $\xi_{60} <0$ from the physical requirement of the increase of entropy for both models. The bound is tighter when the full data set is considered.

For the last two ans\"{a}tze, only upper or lower bounds can be put on the transport coefficients with significantly milder bounds coming from the fit with only late time data set taken into account. Furthermore, from the negative values of the Bayesian factors reported in the tables, it can be concluded that none of the considered cases of our model has a better fit than $\Lambda$CDM to our set of data.

An analogous Hubble function to that considered in this paper has been previously studied in Refs. \cite{Begue:2017lcw, Gao:2021xnk}.
In those papers, the Hubble function arised in the framework of \textit{Quantum Field Cosmology} proposed by Weinberg \cite{Weinberg:2009wa}.
In his framework, the gravitational constant varies with redshift and consequently, the cosmological constant acquires a dependence on the redshift. 

The differences between our results and the precedent findings stem from  different aspects. In the proposed "varying $\Lambda$"CDM model ($\tilde{\Lambda}$CDM),  the scaling of matter and radiation is modified by the same deviation parameter $\delta_G$ that comes from the redshift dependence of the Newton constant $G$, and the deviation parameter for the cosmological constant $\delta_\Lambda$ is related to $\delta_G$ by
\begin{equation}
	\delta_\Lambda \simeq \left( \frac{\Omega_r + \Omega_m}{\Omega_\Lambda} \right) \delta_G
	\label{eq:reldeltaGL}
\end{equation}
for $\delta_\Lambda, \delta_G \ll 1$, with the consistency relation $\delta_G \delta_\Lambda >0$. 

In our model, instead, we have $\delta_m \neq \delta_r$, and the Eq.(\ref{eq:reldeltaGL}) does not hold so that the consistency relation is not compatible with our physical requirements for the deviation parameters, which are not constrained to be much smaller than unity. Our results are also not fully comparable when  $\delta_\Lambda$ and $\delta_G$ are taken as two independent parameters in the so-called extended "varying $\Lambda$"CDM model (e$\tilde{\Lambda}$CDM), with the last parameter still describing the deviation from usual scaling for both matter and radiation. Because of these differences, we have obtained different constraints on our model parameters with respect to those obtained in \cite{Begue:2017lcw, Gao:2021xnk}.

Nevertheless, analogously to what we have found for our model, the two models $\tilde{\Lambda}$CDM and e$\tilde{\Lambda}$CDM, 
 tested against the combined set of CMB distance prior data from  \textit{Planck 2018}, BAO and SNIa Pantheon compilation, partially overlapping our more extended full data set, result to be compatible with $\Lambda$CDM.  Similarly as in our model, these models can not fit the data better than $\Lambda$CDM and cannot relieve the Hubble tension, unless the local measurement by the SH0ES team \cite{Riess:2019cxk} is added to the data set.

\section{ \label{sec:conclusions} Conclusion}
In this paper we have explored for the first time the physical consequences of the recently proposed framework of Ricci Cosmology on the late time Universe. We have derived a simple solution of Ricci Cosmology under the preliminary assumption that the second order transport coefficients are constant. Such an assumption of the non-zero transport coefficients induced some small deviations from perfect fluid scaling of the Universe matter components (dust, radiation, cosmological term). 

Further, the basic constraints (priors) for the parameters made of transport coefficients and describing deviations from the standard cosmology have been obtained from the physical requirement of the increase of entropy according to the second law of thermodynamics. 

Then, we have fitted the full and the late time cosmological data sets reported in section \ref{sec:data} (supernovea, GRBs, cosmic chronometers, H0LiCOW, BAO, CMB), we have found the bounds on the parameters of Ricci cosmology realising that for our simple model it is compatible with standard $\Lambda$CDM cosmology which statistically still fits better to the data. In order to answer the question if Ricci cosmology may give some stronger effects on the evolution of the universe it is perhaps advisable to release the assumption of the constancy of the transport coefficients. This investigation is left for future work.

\section{Acknowledgements}
R.C. would like to thank Rudolf Baier, Sayantani Lahiri and Paul Romatschke for useful discussions.

\appendix

\section{Priors derivation}
\label{sec:appendix}
The general expressions for the deviation parameters in terms of the constant reduced second order transport coefficients are given by
\begin{equation}
    \delta_m = \frac{2 \hat{\xi}_{5m} + \hat{\xi}_{6m}}{1+2 \hat{\xi}_{5m} - \hat{\xi}_{6m}} ,
\end{equation}
\begin{equation}
    \delta_r = \frac{2 \hat{\xi}_{6r}}{1+2 \hat{\xi}_{5r} - \hat{\xi}_{6r}} ,
\end{equation}
\begin{equation}
    \delta_\Lambda = \frac{8 \hat{\xi}_{5 \Lambda} -2 \hat{\xi}_{6 \Lambda}}{1+2 \hat{\xi}_{5 \Lambda} - \hat{\xi}_{6 \Lambda}} .
\end{equation}
Now, we specialize them to the four relevant cases of this paper.

\subsection{Ansatz 1} 
 The assumptions on the reduced second order transport coefficients are 
    \begin{equation}
    \hat{\xi}_{5 \Lambda}  = \hat{\xi}_{5 m}  \equiv \hat{\xi}_{5 0},  \quad \hat{\xi}_{6 \Lambda} = \hat{\xi}_{5 m}  \equiv \hat{\xi}_{6 0} \quad  \hat{\xi}_{5 r}=\hat{\xi}_{6 r} =0 ,
    \end{equation}
    with the deviation parameters given by
    \begin{equation}
    \delta_r = 0, \quad \delta_m = \frac{2 \hat{\xi}_{50} + \hat{\xi}_{60}}{1+2 \hat{\xi}_{50} - \hat{\xi}_{60}} , \quad \textrm{and} \quad
    \delta_\Lambda = \frac{8 \hat{\xi}_{50} -2 \hat{\xi}_{60}}{1+2 \hat{\xi}_{50} - \hat{\xi}_{60}} .
    \label{eq:deltamL}
       \end{equation}
From the entropy increase condition for dust and cosmological constant, we have
\begin{equation}
    \delta_m \leq 0 \quad \textrm{and} \quad 2 \hat{\xi}_{5 0} - \hat{\xi}_{6 0} \geq 0 .
    \label{eq:entropyineq1}
\end{equation}
By using these inequalities in the second of Eq.(\ref{eq:deltamL}), we have
    \begin{equation}
     2 \hat{\xi}_{5 0} + \hat{\xi}_{6 0} \leq 0 .
        \label{eq:ineqxim}
    \end{equation}
By inverting the relations for $\delta_m$ and $\delta_\Lambda$ in Eq.(\ref{eq:deltamL}), we have
\begin{equation}
    \hat{\xi}_{50} = \frac{\delta_m +\frac{1}{2}\delta_\Lambda}{2(3 + \delta_m - \delta_\Lambda)} \quad \textrm{and} \quad   \hat{\xi}_{60} = \frac{ 2 \delta_m -\frac{1}{2}\delta_\Lambda}{3+\delta_m-\delta_\Lambda} .
    \label{eq:ximL}
\end{equation}
By substituting Eq.(\ref{eq:ximL}) in Eq.(\ref{eq:ineqxim}), we have 
\begin{equation}
    \begin{split}
       \frac{3 \delta_m}{3+\delta_m -\delta_\Lambda} \leq 0 ,
    \end{split}
\end{equation}
which together with $\delta_m \leq 0$, implies
\begin{equation}
   \delta_\Lambda < 3+\delta_m .
    \label{eq:ineq1}
\end{equation}
From the second inequality in Eq.(\ref{eq:entropyineq1}), we find that
\begin{equation}
 \frac{ \delta_\Lambda - \delta_m}{3+\delta_m-\delta_\Lambda} \geq 0 ,
\end{equation}
and by using Eq.(\ref{eq:ineq1}), it holds 
\begin{equation}
\delta_\Lambda \geq \delta_m .
    \label{eq:ineq2}
\end{equation}
From inequalities (\ref{eq:ineq1}) and (\ref{eq:ineq2}), we have the following prior for $\delta_\Lambda$
\begin{equation}
    \delta_m \leq \delta_\Lambda <3+\delta_m .
    \label{eq:Lbounds}
\end{equation}  
From the expressions in Eq.(\ref{eq:ximL}) and the last equation we arrive at the following bound for $\hat{\xi}_{6 0}$ 
\begin{equation}
 \hat{\xi}_{6 0} < 0.
\end{equation}
By combining this result with the second inequality in Eq.(\ref{eq:entropyineq1}) and Eq.(\ref{eq:ineqxim}), we have
 \begin{equation}
 -\frac{|\hat{\xi}_{6 0}|}{2} \leq \hat{\xi}_{5 0} \leq \frac{|\hat{\xi}_{6 0}|}{2}
\end{equation}

\subsection{Ansatz 2}
 The assumptions on the reduced second order transport coefficients are
\begin{equation}
 \hat{\xi}_{5 \Lambda} = \hat{\xi}_{5 m} = \hat{\xi}_{5 r} \equiv \hat{\xi}_{5 0},  \quad \hat{\xi}_{6 \Lambda} = \hat{\xi}_{5 m} =\hat{\xi}_{6 r} \equiv \hat{\xi}_{6 0} ,
 \end{equation}
with the deviation parameters given by
\begin{equation}
 \delta_m =  \frac{2 \hat{\xi}_{50} + \hat{\xi}_{60}}{1+2 \hat{\xi}_{50} - \hat{\xi}_{60}}, \quad  \delta_r=  \frac{2 \hat{\xi}_{60}}{1+2 \hat{\xi}_{50} - \hat{\xi}_{60}}  .\label{eq:deltamr}
\end{equation}
 and
\begin{equation}
\delta_\Lambda = \frac{8 \hat{\xi}_{50} -2 \hat{\xi}_{60}}{1+2 \hat{\xi}_{50} - \hat{\xi}_{60}} = 4 \delta_m -3 \delta_r .
\end{equation}
The priors from the increase of entropy are 
\begin{equation}
    2 \hat{\xi}_{5 0} - \hat{\xi}_{6 0} \geq 0, \quad \delta_m\leq 0 \quad \textrm{and} \quad \delta_r \leq 0.
    \label{eq:ineqentropy2}
\end{equation}
From the first two inequalities, it holds that
 \begin{equation}
 2 \hat{\xi}_{5 0} + \hat{\xi}_{6 0} \leq 0 .
\label{eq:ineqxi56}
\end{equation}    
By inverting the expressions in Eq.(\ref{eq:deltamr}), we have
    \begin{equation}
    \hat{\xi}_{50} = \frac{-\delta_r + 2 \delta_m}{4(1-\delta_m +\delta_r)} \quad \textrm{and} \quad   \hat{\xi}_{60} = +\frac{ \delta_r}{2(1-\delta_m +\delta_r)} .
\end{equation}
The inequality (\ref{eq:ineqxi56}) can thus be rewritten in terms of $\delta_m$ and $\delta_r$ as
\begin{equation}
    \frac{\delta_m}{(1-\delta_m +\delta_r)} \leq 0 ,
\end{equation}
which for $\delta_m \leq 0$, gives us a lower bound for $\delta_r$
\begin{equation}
  \delta_r > -1 +\delta_m.
  \label{eq:ineqmrdelta}
\end{equation}
Furthermore, from the first inequality in Eq.(\ref{eq:ineqentropy2}), and the lower bound in Eq.(\ref{eq:ineqmrdelta}), we obtain
\begin{equation}
    \delta_r \leq \delta_m .
\end{equation}
Thus, the physical bounds on the deviation parameter $\delta_r$ is given by
\begin{equation}
    -1 +\delta_m < \delta_r \leq \delta_m.
\end{equation}
For the reduced transport coefficients, from the first and third inequalities in Eq.(\ref{eq:ineqentropy2}), we arrive at  \begin{equation}
\hat{\xi}_{6 0} \leq 0 ,
\end{equation}
which combined with the first inequality in Eq.(\ref{eq:ineqentropy2}) and Eq.(\ref{eq:ineqxi56}), finally gives us \begin{equation}
-\frac{|\hat{\xi}_{6 0}|}{2} \leq \hat{\xi}_{5 0} \leq \frac{|\hat{\xi}_{6 0}|}{2} .
\end{equation}
   
\subsection{Ansatz 3} 
The assumptions on the reduced second order transport coefficients are
    \begin{equation}
        \hat{\xi}_{5 \Lambda} =\hat{\xi}_{6 \Lambda}, \quad \hat{\xi}_{5 m} =\hat{\xi}_{6 m}, \quad \textrm{and} \quad ,
        \hat{\xi}_{5 r} =\hat{\xi}_{6 r} = 0
    \end{equation}
    with the deviation parameters that read
    \begin{equation}
        \delta_m = \frac{3 \hat{\xi}_{5 m}}{1+\hat{\xi}_{5 m}} \quad \textrm{and} \quad \delta_\Lambda =  \frac{6 \hat{\xi}_{5 \Lambda}}{1+\hat{\xi}_{5 \Lambda}} .
    \end{equation}
The priors from the increase of entropy are given by
\begin{equation}
    \delta_m \leq 0, \quad \textrm{and} \quad \hat{\xi}_{5 \Lambda} \geq 0,
\end{equation}
which imply the following bounds for the deviation parameter$\delta_\Lambda$ and reduced transport coefficient $\hat{\xi}_{5 m}$
\begin{equation}
   -1 <\hat{\xi}_{5 m} < 0, \quad \textrm{and} \quad \delta_\Lambda \geq 0 .
\end{equation}

    \subsection{Ansatz 4}
The assumptions on the reduced second order transport coefficients are
\begin{equation}
 \hat{\xi}_{5 \Lambda} =\hat{\xi}_{6 \Lambda}, \quad \hat{\xi}_{5 m} =\hat{\xi}_{6 m}, \quad \textrm{and} \quad \hat{\xi}_{5 r} =\hat{\xi}_{6 r}
\end{equation}
with the deviation parameters that read 
\begin{equation}
\delta_m =  \frac{3 \hat{\xi}_{5 m}}{1+\hat{\xi}_{5 m}}, \quad \delta_\Lambda = \frac{6 \hat{\xi}_{5 \Lambda}}{1+\hat{\xi}_{5 \Lambda}} \quad  \textrm{and} \quad \delta_r = \frac{2 \hat{\xi}_{5 r}}{1+\hat{\xi}_{5 r}} .
\end{equation}
The priors from the increase of entropy are given by \begin{equation}
    \delta_r \leq 0,  \quad \delta_m \leq 0, \quad \textrm{and} \quad \hat{\xi}_{5 \Lambda} \geq 0,
\end{equation}
which imply the following bounds for the deviation parameter$\delta_\Lambda$ and reduced transport coefficients $\hat{\xi}_{5 r}$ and $\hat{\xi}_{5 m}$ 
\begin{equation}
  -1 <\hat{\xi}_{5 r} < 0, \quad -1 <\hat{\xi}_{5 m} < 0, \quad \textrm{and} \quad \delta_\Lambda \geq 0 .
\end{equation}

\bibliography{01072021_Ricci_Paper_1}

\end{document}